\date{}
\documentclass[aps,pra,showpacs]{revtex4-1}
\usepackage{graphicx,psfrag,amsmath,amssymb,amsfonts,latexsym,color,dcolumn}

\newcommand{\pp}{\mathbf{p_{\parallel}}}
\newcommand{\mpp}{\vert\mathbf{p_{\parallel}}\vert}

\begin{document}
\title{A functional approach to quantum friction: effective
action and dissipative force}

\author{M. Bel\' en Far\'{\i}as $^1$~\footnote{mbelfarias@df.uba.ar}}
\author{C\'esar D. Fosco $^2$}
\author{Fernando C. Lombardo$^1$}
\author{Francisco D. Mazzitelli$^2$}
\author{Adri\'an E. Rubio L\'opez$^1$}

\affiliation{$^1$ Departamento de F\'\i sica {\it Juan Jos\'e
Giambiagi}, FCEyN UBA and IFIBA CONICET-UBA, Facultad de Ciencias Exactas y Naturales,
Ciudad Universitaria, Pabell\' on I, 1428 Buenos Aires, Argentina}
\affiliation{$^2$ Centro At\'omico Bariloche and Instituto Balseiro,
Comisi\'on Nacional de Energ\'\i a At\'omica, 8400 Bariloche,
Argentina}

\date{today}
\begin{abstract}  
We study the Casimir friction due to the relative, uniform, 
lateral motion of two parallel semitransparent mirrors coupled to a vacuum real scalar
field, $\phi$.  We follow a functional approach, whereby nonlocal terms in the
action for $\phi$, concentrated on the mirrors' locii,
appear after functional integration of the microscopic degrees of freedom.
This action for $\phi$, which incorporates the relevant properties of the
mirrors, is then used as the starting point for two
complementary evaluations: Firstly, we calculate the  { in-out}
effective action for the system, which develops an imaginary part, hence  a
non-vanishing probability for the decay (because of friction) of the
 initial vacuum state.  Secondly, we evaluate another observable: the 
vacuum expectation value of the frictional force, using the { in-in} or Closed Time Path
formalism.  Explicit results are presented for zero-width mirrors and half-spaces, in a
model where the microscopic degrees of freedom at the mirrors are a set of
identical quantum harmonic oscillators, linearly coupled to $\phi$. 
\end{abstract}
\maketitle
\section{Introduction}\label{sec:intro} 
The quantum nature of microscopic systems may, under some special
circumstances, manifest itself in the form of interesting macroscopic
effects. On the other hand, one of the most distinctive features of
quantum phenomena are the vacuum fluctuations, among which one of the most
celebrated examples are the zero-point electromagnetic field
fluctuations.  These, however, do not produce any observable effect in free
space. This may change drastically when non-trivial boundary conditions are
imposed on the electromagnetic field: in the Casimir effect (and related phenomena), a
force appears even between two {\em neutral\/} macroscopic bodies. Indeed, this
effect can be understood as due to the fact that vacuum fluctuations induce
(vacuum) currents in each object, the interaction between which results in
a macroscopic force~\cite{libros}.

The very same quantum fluctuations may also produce qualitatively different
observable effects in different set-ups. One that has received much
attention is the existence of a frictional force when plane mirrors {\em
which are not in contact\/} undergo constant-speed relative parallel
motion.  
 At the classical level, a distribution of electrical charges outside a dielectric surface induces image 
charges of the opposite sign, that produce an attractive force on the 
external charge distribution. If the external charges move parallel to the surface, for lossy media
the position of the images does not coincide with  the instantaneous specular reflection, giving rise to a lateral, frictional
force. At the quantum level, in the case of  of two  flat parallel mirrors 
separated by vacuum, the zero-point energy of the electrons on each surface
produces charge fluctuations, that in turn  induce image charges 
on the other, giving rise to the static Casimir force \cite{libros}. When lossy mirrors are set in relative parallel motion, a frictional force 
is generated by the phase
lag between the charges and currents induced by the vacuum fluctuations on each surface.
That phase lag is not
present for perfect mirrors~\cite{Pendry97,Pendry2010}. A different situation, which
also leads to friction, is due to the quantum Cerenkov effect between
non-dispersive media~\cite{Maghrebi:2013jpa} at a relative speed which
surpasses a threshold determined by the speed of light in the media.  In
any case, the effect can be understood in terms of an exchange of virtual
photons between two bodies, which in turn excite their internal degrees of
freedom.  This effect has been analyzed~\cite{Pendry97} (and
debated~\cite{Pendry2010,debate}) at length, mainly for the case of media which fill
half-spaces, shifting with constant velocity. The frictional force
between two atoms in relative constant motion has also been computed, along with 
the dissipative force acting on an atom moving parallel to a plate with constant 
velocty: in Ref.\cite{others}, these geometries are studied using microscopic simple models for the atoms.
Ref.\cite{vp2007} contains a detailed account of the works
on the subject,  pointing out
some contradictory results in earlier literature.  
Note that quantum
dissipative effects on moving bodies may also be due to the excitation of
{\em real\/} photons out of the quantum vacuum, an effect known as
dynamical Casimir effect (see, for instance,
Ref.~\cite{dce}). The latter, however, unlike
the quantum friction phenomenon, requires the existence on non-vanishing
accelerations.  

In this paper, we present a detailed study of quantum friction between two
mirrors which undergo constant parallel speed relative motion, using  functional
methods. We follow
two complementary approaches that, we believe, shed new light on this
interesting effect from the perspective of quantum field theory. We present our study for a specific simple model, consisting of a vacuum
scalar field linearly coupled to a set of uncoupled quantum harmonic
oscillators which are the microscopic `matter' degrees of freedom on the mirrors. This is the simpler way of modelling 
microscopic degrees of freedom,
and has been used\cite{others} to model atoms and calculate the frictional force between them. 

Our first approach here is analogous to the one presented in a previous
paper by some of us~\cite{Fosco2011}, where dissipative effects (for either
normal or parallel motions) have been analysed using an Euclidean
functional integral formalism for the calculation of the effective action,
the result of which is rotated back to real time.  Mirrors have been
represented by nonlocal coupling terms in the vacuum field action, which
proceed from the integration of the microscopic degrees of freedom.  It has
been shown there that, indeed, an imaginary part for the in-out effective
action emerged as a consequence of non-contact friction. We extend here that
study in more than one direction: we present a more detailed
description of the model for the microscopic degrees of freedom (namely,
before integrating them out) and we analyze in detail the relation
between the analytic structure of the nonlocal coupling terms in Fourier
space and the presence of frictional forces, performing all the calculations in
real time, and discussing the subtleties of Wick rotation.   The second approach 
consists in the explicit computation of the frictional force from the vacuum expectation value of the energy momentum tensor.
 We will see that, although the problem is stationary, as the {\it in} and {\it out} vacuum states 
 of the system do not coincide, it is necessary to use the  {in-in} or Closed Time Path (CTP) formalism.
Both the { in-out} and {in-in}  functional approaches have been previously applied
to the case of accelerated mirrors in Ref.\cite{Fosco:2007nz}. 

Regarding the mirrors, we consider two different geometries: two infinitesimally
thin mirrors (that is, two-dimensional mirrors of zero width) separated by
a distance $a$, and two half-spaces separated by a distance $a$. 

The structure of this paper is as follows: in Section~\ref{sec:system} we
define the class of system that we consider in this paper and establish
some of the approximations to be used.  We also present a microscopic model which provides concrete
realizations of the system defined above.  Then, in
Section~\ref{sec:effective}, we calculate the in-out effective action,
studying the relation between its analytic structure and the existence of
friction.  In Section~\ref{sec:force} we calculate the frictional force, by
means of the in-in vacuum expectation value of the stress tensor, using the CTP formalism.
Section~\ref{sec:concl} contains our conclusions.

\section{The system}\label{sec:system}
Let us begin by defining the (real-time) action ${\mathcal S}$ for the
system; it depends on the vacuum field $\phi$ and on the matter
fields, denoted collectively by $\psi$, confined to the mirrors. 
Hence, the action naturally decomposes into three terms, as follows: 
\begin{equation}\label{eq:defs}
{\mathcal S}[\phi, \psi] \;=\; {\mathcal S}^{(0)}_{\rm v}[\phi] 
\,+\, {\mathcal S}^{(0)}_{\rm m}[\psi] \,+\, {\mathcal S}^{({\rm int})}_{\rm v m}[\phi,\psi] , 
\end{equation}
where ${\mathcal S}^{(0)}_{\rm v}$ is the free (i.e., no mirrors) action for the vacuum
field:
\begin{equation}\label{eq:defs0}
{\mathcal S}^{(0)}_{\rm v}[\phi]\;=\; \frac{1}{2} \int d^4x \,\big[ \partial^\mu
\phi\partial_\mu \phi \,-\, (m^2-i \epsilon) \phi^2 \big] \;,
\end{equation}
whilst ${\mathcal S}^{(0)}_{\rm m}$ and ${\mathcal S}_{\rm v m}^{({\rm int})}$ denote the actions for the free
matter field and for the  $\phi -\psi$ interaction, respectively.
Assuming locality of the microscopic vacuum-field/matter interaction,  
${\mathcal S}_{\rm v m}^{({\rm int})}$ will only depend on the vacuum field at spatial points
on the regions occupied by the two mirrors, which we will denote by $L$ and
$R$ (each letters will be used to denote both a mirror and the spatial
region it occupies). Each mirror is assumed to have homogeneous and
isotropic properties on each $x^3 = {\rm constant}$ plane, whenever $x^3$
is inside the mirror width. Besides, those properties are independent of
$x^3$ inside each mirror. We adopt the convention $\hbar=c=1$.

The in-out effective action $\Gamma$ for the full system described by
${\mathcal S}$ may therefore be written in terms of  the vacuum persistence
amplitude, ${\mathcal Z}$, which in turn can be represented as a functional
integral: 
\begin{equation}\label{eq:funcgenphipsi}
e^{ i \Gamma} \;=\; \mathcal{Z}=\;\langle0_{\rm out}\vert 0_{\rm in}\rangle\;=\int \mathcal{D}\phi \mathcal{D}\psi \;
e^{i {\mathcal S}[\phi,\psi]}\;. 
\end{equation}

Rather than performing the functional integrals over matter and vacuum
fields simultaneously, it is convenient to introduce the partial result of
integrating out just the matter degrees of freedom at the plates:
\begin{equation}\label{eq:equrep1}
\mathcal{Z} \;=\;\int \mathcal{D}\phi \; e^{i {\mathcal S}_{\rm v}^{({\rm eff})}[\phi]}\;, 
\end{equation}
with ${\mathcal S}_{\rm v}^{({\rm eff})}[\phi] \equiv {\mathcal S}^{(0)}_{\rm v}[\phi]
+ {\mathcal
S}^{({\rm int})}_{\rm v} [\phi]$, where the second term incorporates the effect of the
matter degrees of freedom, and is given by 
\begin{equation}\label{eq:defSvm}
e^{i {\mathcal S}^{({\rm int})}_{\rm v}[\phi]}\;=\; \int \mathcal{D}\psi \; 
e^{i \big( {\mathcal S}^{(0)}_{\rm m}[\psi] + {\mathcal S}_{\rm v m}^{({\rm
int})}[\phi,\psi]\big)}\;. 
\end{equation}

Regardless of the model used for the mirrors, based on the assumptions
about the system, the general form of ${\mathcal S}^{({\rm int})}_{\rm v}[\phi]$
will be:
\begin{equation}
{\mathcal S}^{({\rm int})}_{\rm v}[\phi]\;=\; {\mathcal S}^{(L)}_{\rm v}[\phi] +
{\mathcal S}^{(R)}_{\rm v}[\phi]\;, 
\end{equation}
where ${\mathcal S}^{(R)}_{\rm v}[\phi]$ and  ${\mathcal S}^{(L)}_{\rm v}[\phi]$ are,
in general, nonlocal and non-quadratic functionals of
$\phi(x_\parallel,x^3)$, where $x_\parallel \equiv (x^0,x^1,x^2)$. Because
of the assumed locality of the microscopic interaction, we also know that
${\mathcal S}^{(L,R)}_{\rm v}$ will depend on $\phi(x_\parallel,x^3)$ only for $x^3$
inside the region defining the respective mirror. 
It is convenient to introduce, in this respect, two functions $\chi_L(x^3)$ and
$\chi_R(x^3)$, respectively, which determine the regions occupied by them.
For finite or infinite width mirrors: $\chi_{L,R}(x^3) = 1$ if $x_3$ belongs to
$L,R$, and   $\chi_{L,R}(x^3) = 0$ otherwise. For zero-width ones, they are
just $\delta$ functions of the corresponding value of $x^3$.

Thus, under the assumption that, either exactly (as in the model we shall
consider) or approximately, ${\mathcal S}^{({L,R})}_{\rm v}$ is quadratic in its
respective argument, we
have:
\begin{equation}
{\mathcal S}^{({\rm int})}_{\rm v}[\phi] \;=\; - \frac{1}{2} 
\int_{x,y} \phi(x) V(x,y) \phi(y) 
\end{equation}
(where we introduced a shorthand notation for the two spacetime integrals) with:
\begin{equation}
V(x,y) \;=\; V_L(x,y) \,+\, V_R(x,y) \;.
\end{equation}
and 
\begin{equation}
V_{L,R}(x,y) \;=\; \chi_{L,R}(x^3) \, \delta(x^3-y^3) \, \lambda_{L,
R}(x_\parallel-y_\parallel) \;.
\end{equation}
The `potentials' $V_{L,R}$ can be determined by using a concrete model,
or even introduced by hand, under some specific assumptions. 
Nevertheless, regardless of the origin of those potentials, the $\phi$
integral becomes a Gaussian,
\begin{equation}
\label{eq:phigaussian}
{\mathcal Z}\;=\;\int \mathcal{D}\phi \, e^{-\frac{1}{2} \int_{x,y}\,
\phi(x) A(x,y) \phi(y) } \;,
\end{equation}
where we introduced $A(x,y)$, which may be regarded as the kernel of an
(integral) operator $A$. In a Dirac bracket-like notation: \mbox{$A(x,y) =
\langle x|A|y\rangle$}, with:
\begin{equation}\label{eq:defA}
A(x,y)\;=\; \big[ i (\Box_x + m^2 ) + \epsilon \big] \delta(x-y) + i V(x,y) \;.
\end{equation}

Thus, the formal result of the integral over $\phi$ yields for $\Gamma$:
\begin{equation}
\label{accionefectiva}
\Gamma\;=\;\frac{i}{2} \, {\rm Tr}\log A \,.
\end{equation}
An expansion of $\Gamma$ in powers of the potentials can be performed by
noting that $A = A_0 + A_1$, where 
\begin{equation}\label{eq:defA0}
A_0(x,y)\;=\; \big[ i (\Box_x + m^2 ) + \epsilon \big] \delta(x-y)  \;,
\end{equation}
is the inverse of the free Feynman propagator 
$G_F(x-y)=-i$\mbox{$\langle 0|T[\phi(x)\phi(y)|0\rangle$}, and 
\begin{equation}\label{eq:defA1}
A_1(x,y)\;=\; i V(x,y) \;.
\end{equation}
The first contribution in this expansion which already encodes a nontrivial
interaction between the two mirrors is of the second order, and has the form:
\begin{equation}\label{eq:gammaI}
\Gamma_I^{(2)} \;=\; -\frac{i}{2} {\rm Tr} \big(G_F V_L G_F V_R\big) \;.
\end{equation}
The trace may be evaluated in momentum space, so that
\begin{equation}\label{eq:effam}
\Gamma_I^{(2)} \;=\;- \frac{i}{2} \int \frac{d^4p}{(2\pi)^4}
\frac{d^4q}{(2\pi)^4} \tilde{G}_F(p) \tilde{G}_F (q) \tilde{V}_L(p,q)
\tilde{V}_R(q,p) \;,
\end{equation}
where $\tilde{G}(p) \equiv \frac{i}{p^2 - m^2 + i \epsilon}$, while the two momentum space kernels $\tilde V_{R,L}$ are determined by the geometry and composition of the mirrors, as well as by the relative motion between them. The advantage of using a microscopic model is that the analytic properties of the kernels will be completely determined after the integration of the matter degrees of freedom.
Let us consider now, in the next subsection, how the effect of the relative
motion is reflected in the potentials.
\subsection{Potentials}\label{sec:pots}
Since only the relative motion of the mirrors may affect the physical
results, we shall use as the reference system a laboratory frame ($L$), where $L$
is at rest,  while $R$ moves rigidly with a constant speed $u$ along any direction
parallel to its homogeneity and isotropy planes, $x^1$ say. 

Using $x'^\mu$, $\mu = 0,1,2,3$ for coordinates fixed to the moving mirror,
and assuming $|u| << 1$, we have the Galilean transformations:
$x^0 = x'^0$, $x'^0 = x^0$, $x'^1 = x^1 - u x^0$, $x'^2 = x^2$ and $x'^3 =
x^3$.

For the $L$ mirror, under the assumptions we presented above, the potential
necessarily has the form:
\begin{equation}\label{eq:vllab}
V_L(x,y)\;=\; \chi_L(x^3) \,\lambda_L(x_\parallel-y_\parallel) \,
\delta(x^3-y^3) \;, 
\end{equation}
where $\lambda_L$ may be conveniently determined by its Fourier space
transformed $\tilde\lambda_L(k^0,k^1,k^2)$. 
Regarding the $R$ mirror, we note that, in a comoving reference system, 
\begin{equation}\label{eq:vrcm}
V'_R(x',y')\;=\; \chi_R(x'^3) \,\lambda_R(x'_\parallel-y'_\parallel) \,
\delta(x'^3-y'^3) \;, 
\end{equation}
where $\lambda_R$ is determined by the microscopic model at rest, since it
is a comoving system object. It may be
written in terms of its Fourier transform, as follows:
\begin{equation}
	\lambda_R(x'_\parallel-y'_\parallel)\;=\; \int
	\frac{d^3k_\parallel}{(2\pi)^3} \, e^{i k_\parallel \cdot
	(x'_\parallel - y'_\parallel)}
	\tilde\lambda_R(k_\parallel)
\end{equation}
(no need to introduce primed variables for the momenta, since they are
integrated, dummy variables). Note that, if the two media were identical,
$\tilde\lambda_R$ above would be identical to
$\tilde\lambda_L$.

The interaction between the mirrors and the vacuum field will be:
\begin{equation}
	{\mathcal S}_{\rm v}^{\rm (int)} \;=\; - \frac{1}{2} \int_{x, y} \, \phi(x)
	\, V_L(x,y) \phi(y) \,
- \, \frac{1}{2} \int_{x',y'}\, \phi'(x') \, V'_R(x',y') \phi'(y') \;.
\end{equation}
We have to put both potentials in the same reference system. The scalar
field satisfies $\phi'(x') = \phi(x)$, and $\chi_R$ is invariant. On the
other hand:
\begin{eqnarray}
\lambda_R(x'_\parallel-y'_\parallel) &=& \int
	\frac{d^3k_\parallel}{(2\pi)^3} \, e^{i [k^0 (x'^0 - y'^0) - 
k^1 (x'^1 - y'^1) -  k^2 (x'^2 - y'^2)}
\tilde\lambda_R(k^0,k^1,k^2) \nonumber\\
&=& \int \frac{d^3k_\parallel}{(2\pi)^3} \, e^{i [k^0 (x^0 - y^0) - 
k^1 (x^1 - y^1 - u (x^0 - y^0)) -  k^2 (x^2 - y^2)]}
\tilde\lambda_R(k^0,k^1,k^2) \nonumber\\
&=& \int \frac{d^3k_\parallel}{(2\pi)^3} \, e^{i [k^0 (x^0 - y^0) - 
k^1 (x^1 - y^1) -  k^2 (x^2 - y^2)]}
\tilde\lambda_R(k^0- u k^1,k^1,k^2) \;,
\end{eqnarray}
where we have used the Galilean transformation and a shift of integration
variables. The last line in the equation above tells us that, in the $L$
system, the $R$ mirror is described by the shifted $\tilde\lambda_R$
function:
\begin{equation}
\tilde\lambda_R(k^0,k^1,k^2) \;\to\;
\tilde\lambda_R(k^0- u k^1,k^1,k^2)\;.
\end{equation} 
\subsection{Microscopic model for the media}\label{ssec:micro}
We introduce here a simple microscopic model, a concrete
realization of the interaction between vacuum and matter fields, which
provides a physically acceptable function $\tilde{\lambda}$.
Microscopic matter degrees of freedom on the media behave as
one-dimensional harmonic oscillators, one at each point of the mirror. 
They have generalized coordinates $Q(x^0,x^1,x^2) = Q(x_\parallel)$, taking
values in an internal space.
No coupling between the oscillators is included, and there is a linear
coupling between each oscillator and the vacuum field.  The interaction
only occurs locally, at the spatial positions occupied by the media. 

To find ${\mathcal S}_{\rm v}^{(L)}$, we consider the terms in the action
depending on $Q_L$ (for $R$ an analogous argument will apply): 
\begin{equation}
	{\mathcal S}_{\rm m}^{(0)} \;=\; \frac{1}{2}\int d^4x \chi_L(x^3)\,
	\big[\dot{Q}_L^2(x_\parallel)-(\Omega_L^2-i \epsilon)
Q_L^2(x_\parallel)\big] 
\end{equation}
and
\begin{equation}
{\mathcal S}^{\rm (int)}_{\rm v m} \;=\;  g_L \int d^4x \, \chi_L(x^3) \,
Q_L(x_\parallel) \phi(x) \;.
\end{equation}
The integral to find ${\mathcal S}_{\rm v}^{(L)}$ is a Gaussian, and it results
in the potential:
\begin{equation}
	V_L(x,y) \;=\;\chi_L(x^3) \delta(x^3-y^3)\,\lambda_L(x_\parallel-y_\parallel)
\end{equation}
with the Fourier transform of $\lambda_L$ given by:
\begin{equation}
	\tilde\lambda_L(k_\parallel) \;=\;
	{\tilde\lambda}_L(k^0) \;,
\end{equation}
with
\begin{equation}\label{eq:lambda}
{\tilde\lambda}_L(k^0) \; = \,
\frac{g_L^2}{(k^0)^2-\Omega_L^2+i\epsilon}\;.
\end{equation}
 
Note that, even for this simple model, $\tilde{\lambda}_L$ is
not analytic, since it has two poles, located at $k^0_L=\pm\sqrt{\Omega_L^2-i\epsilon}\approx
\pm \Omega_L \mp \frac{i\epsilon}{2\Omega_L}$.

An important remark is in order: the mass dimensions of the coupling constant
$g_L$ and of $Q$ are different when $\chi_L$ is a $\delta$ function rather
than a step-like function. Indeed, in the former, $[Q] = 1/2$ and
$[g_L]=-3/2$, while in the latter $[Q] = 1$ and
$[g_L]=-2$.
For a moving $R$ mirror, on the other hand, we shall have
\begin{equation}
{\tilde\lambda}_R(k^0,k^1) \; = \, \frac{g_R^2}{(k^0 - u
k^1)^2-\Omega_R^2+i\epsilon} \;.
\end{equation}

\section{In-out effective action}\label{sec:effective}
Let us now compute the effective action as a function
of the $\tilde{\lambda}$ functions which characterize the material:
\begin{equation}
	\Gamma_I^{(2)}\;=\; \frac{-iT \Sigma}{2(2\pi)^2} \int
	d^3p_\parallel \; \tilde\lambda_L(p^0)
	\tilde\lambda_R(p^0-u p^1) \;
	\int dx^3 dy^3 \; \chi_L(x^3) \,
	\big[G(p_\parallel,x^3-y^3)\big]^2 \, \chi_R(y^3) \, ,
\end{equation}
where $T$ is the total time, $\Sigma$ the total surface of the
plates, and:
\begin{equation}
	G(p_\parallel,x^3) \;=\; i \, \int dp^3 \, \frac{e^{i p^3
	x^3}}{(p_\parallel)^2 - (p^3)^2 + i \epsilon} \;. 
\end{equation}

For two zero-width mirrors, at a distance $a$, we obtain:
\begin{equation}
\label{eq:accionef}
\Gamma_I^{(2)}\;=\; \frac{i T \Sigma}{4} \int d^3 p_\parallel \,
\frac{e^{2ia\sqrt{(p_\parallel)^2+i\epsilon}}}{(p_\parallel)^2+i\epsilon}
\;\tilde\lambda_L(p^0) \, \tilde\lambda_R(p^0-u p^1)\;.
\end{equation}
Here, each  $\tilde{\lambda}(\omega)$ appears evaluated at a frequency
measured at the rest frame of each plate.

On the other hand, for infinite media filling half-spaces, namely,
$\chi_L(x^3) = \theta(-x^3)$ and $\chi_R(x^3) = \theta(x^3-a)$, we see that
\begin{equation}
\label{eq:accionefectivasemiespacios}
\Gamma_I^{(2)}\;=\; \frac{i T \Sigma}{4} \int d^3 p_\parallel \,
\frac{\tilde\lambda_L(p^0) \; \tilde\lambda_R(p^0- u
p^1)}{(p_\parallel)^2+i\epsilon} \int dx_3 dx'_3 \theta(-x_3)
\theta(x_3'-a) e^{2i(x_3'-x_3)\sqrt{(p_\parallel)^2+i\epsilon}} \;.
\end{equation}

We note that the result corresponding to the two half-spaces may also be
obtained from the one corresponding to thin mirrors, by performing
integrations over two auxiliary variables. Indeed, using the relations:
\begin{equation}
	\theta(-x^3) \;=\; \int_{-\infty}^0 ds_L \,\delta(x^3 - s_L)
	\;,\;\;
	\theta(x^3 -a) \;=\; \int_a^\infty ds_R \,\delta(x^3 - s_R) \;,
\end{equation}
and that $\Gamma$, for thin mirrors, is only a function of the distance
between the mirrors, we obtain:
\begin{equation}\label{eq:hat-trick}
\Gamma_I^{(2)}\;=\; \frac{i T \Sigma}{4} \int_{-\infty}^0 ds_L
\int_{a-s_L}^\infty ds_R
\int d^3 p_\parallel \,
\frac{e^{2i s_R\sqrt{(p_\parallel)^2+i\epsilon}}}{(p_\parallel)^2+i\epsilon}
\;\tilde\lambda_L(p^0) \, \tilde\lambda_R(p^0-u p^1)\;,
\end{equation}
or
\begin{eqnarray}\label{eq:hat-trick2}
\Gamma_I^{(2)} &=& \frac{i T \Sigma}{4} \int_a^\infty ds_R
\int_{a-S_R}^0 ds_L \int d^3 p_\parallel \,
\frac{e^{2i s_R\sqrt{(p_\parallel)^2+i\epsilon}}}{(p_\parallel)^2+i\epsilon}
\;\tilde\lambda_L(p^0) \, \tilde\lambda_R(p^0-u p^1) \nonumber\\
&=& \frac{i T \Sigma}{4} \int_a^\infty ds_R (s_R -a)  \int d^3 p_\parallel \,
\frac{e^{2i s_R\sqrt{(p_\parallel)^2+i\epsilon}}}{(p_\parallel)^2+i\epsilon}
\;\tilde\lambda_L(p^0) \, \tilde\lambda_R(p^0-u p^1) \, .
\end{eqnarray}
Since the auxiliary variable is real, one can also extract the imaginary
part of $\Gamma$ for the half-spaces from the result corresponding to thin
mirrors. Therefore, in what follows we will describe in detail the calculations for the case of thin mirrors,
and eventually quote only the final results for half-spaces. To simplify the notation, we will also omit
the superscript in the second order approximation to the effective action, that will be denoted by
$\Gamma_I$.

\subsection{Imaginary part of the in-out effective action}
Since the in-out effective action  is related to the vacuum persistence amplitude
(Eq.~\eqref{eq:funcgenphipsi}), the presence of an imaginary part signals,
for the systems considered in this paper, the excitation of internal
degrees of freedom on the mirrors. Since this is due to the
constant-velocity motion of one of the mirrors, it reflects the existence
of non-contact friction. 
In this section we will obtain explicit expressions
for this imaginary part, for the microscopic model described above, in the case of zero-width mirrors. The case of media filling
half-spaces will be considered at the end, taking advantage of the result for thin mirrors.

In what follows, we consider the case of identical mirrors, so we shall
drop the $L$ and $R$ subscripts from the microscopic model parameters.
If we use the notation  $\pp=(p^1,p^2)$, the integrand for the effective action of Eq.\eqref{eq:accionef}, considered as a function of
$p^0$, has singularities in $\pm \sqrt{\pp^2-i\epsilon} \approx
\pm (\mpp - i \epsilon / 2 \mpp)$. It also has two branch
cuts: the first one could be taken as starting
on the first singularity, to $+\infty$, parallel to the $x$ axis, (that is, with
$\text{Im}(p^0)=-\epsilon/2 \mpp$ and $\text{Re}(p^0) >
\mpp$, under the approximation of small $\epsilon$). The other
branch cut extends parallel to the real axis, from the second singularity to
$-\infty$.

In Eq.(\ref{eq:accionef}), we can write the integral in the variable $p^0$ in the positive axis
\begin{equation}
\label{primercuad}
\Gamma_I=\frac{iT\Sigma}{4}\int d^2\pp \int_{0}^{\infty}dp^0(f(p^0)+f(-p^0))\, ,
\end{equation}
where
\begin{equation}
	f(p^0)=\frac{e^{2ia\sqrt{(p^0)^2-\pp^2+i\epsilon}}}{(p^0)^2-\pp^2+i\epsilon}\tilde{\lambda}(p^0)\tilde{\lambda}(p^0-u
	p^1) .  
\end{equation}
This allows us to compute the $p^0$-integral in the complex plane by considering 
a closed contour formed by the positive real and imaginary axes and a quarter of a
circle with very large radius. As the integral over the quarter of circle vanishes when the radius of the 
circle tends to infinity, the integral in Eq.(\ref{primercuad}) is given by its Wick rotated
expression $p^0\to i p^0$ plus the contribution coming from the poles of $f(p^0)$ in the first 
quadrant.

As a first example, one may consider the case of constant
$\tilde{\lambda}(p^0)$.  Denoting this constant by  $\omega_{\rm p}^2$, 
it can be shown that this corresponds to the
dielectric permittivity given by the plasma model 
$\epsilon(p^0)=1-\omega_{\rm p}^2/(p^0)^2$. In this case, as $\tilde\lambda$ has no poles,
and $f(i p^0)$ is a real function, there is no imaginary part in the effective action
and therefore no quantum friction.

Let us now consider the case of the microscopic model with uncoupled harmonic oscillators. The function $f(p^0)$ reads
\begin{equation}
\label{eq:accionefosc1}
f(p^0)=g^4 \frac{1}{(p^0)^2-\Omega^2+i\epsilon} \times 
\frac{1}{(p^0-u p^1)^2-\Omega^2+i\epsilon}
\times
\frac{e^{2ia\sqrt{(p^0)^2-\pp^2+i\epsilon}}}{(p^0)^2-\pp^2+i\epsilon} \, , 
\end{equation}
and it has, in addition to the already mentioned singularities, four simple poles, located at:
\begin{align*}
p^0 &=  \sqrt{\Omega^2-i\epsilon} \approx \Omega - \frac{i\epsilon}{2\Omega}\\
p^0 &= -  \sqrt{\Omega^2-i\epsilon} \approx - \Omega + \frac{i\epsilon}{2\Omega}\\
p^0 &=  u p^1 +  \sqrt{\Omega^2-i\epsilon} \approx u p^1 + \Omega - \frac{i\epsilon}{2\Omega}\\
p^0 &=  u p^1 - \sqrt{\Omega^2-i\epsilon} \approx u p^1 -\Omega + \frac{i\epsilon}{2\Omega} \, .
\end{align*}
\begin{figure}[h]
\includegraphics[scale=0.6]{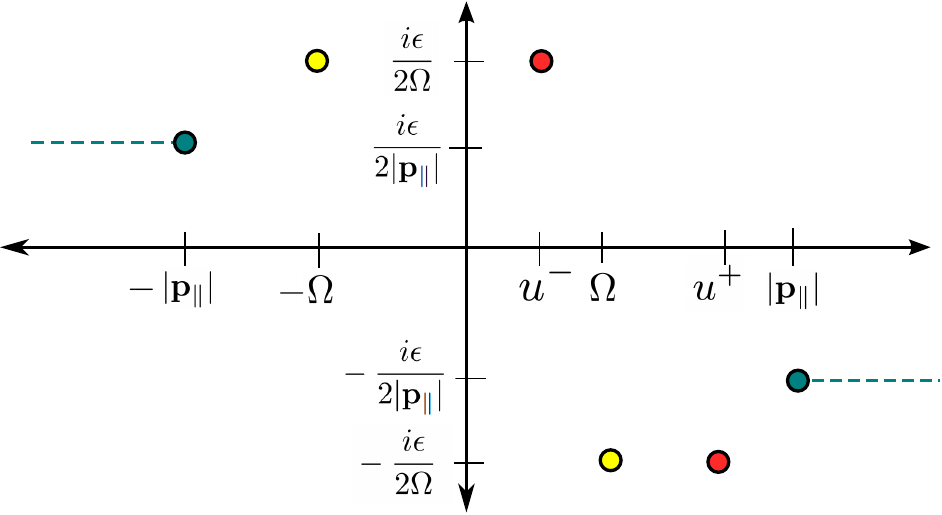}
\caption{\label{fig:polos} (Color online). Singularities of $f(p^0)$ (Eq.(38)) in the complex $p^0$ plane. Simple poles are depicted as filled dots, while the branch cuts are represented by dashed lines. We have introduced the notation: $u^\pm = u p^1 \pm \Omega$.}
\end{figure}

The singularities of $f(p_0)$ are shown in Fig.~\ref{fig:polos} (the
ones for $f(-p^0)$ can be found by $p^0 \rightarrow -p^0$). Note that,
for each term of the integrand ($f(p^0)$ and $f(-p^0)$), there is only one
pole that may appear in the first quadrant,  as long as the parameters fulfill
certain conditions.  For the first term, this happens if $u p^1 - \Omega >
0$ while for the second term when $-u p^1 - \Omega > 0$. 
Then, using Cauchy's theorem we find
\begin{align}
\label{eq:accionefectivaenfunciondep1}
\Gamma_I &= \frac{i T \Sigma}{4} \int d^2\pp\int_{0}^{+\infty} dp^0 (f(p^0)+f(-p^0))= \frac{- T \Sigma}{4} \int d^2 \textbf{p}_\parallel \left\lbrace \int_{0}^{+\infty} dp^0 (f(ip^0)+f(-ip^0)) + \right. \nonumber \\ 
&\left. + \Theta(u p^1 - \Omega)  2 \pi   \text{Res}\left[f(p^0), u p^1 - \sqrt{\Omega^2-i\epsilon}\right] + \Theta(-u p^1 - \Omega) 2 \pi   \text{Res}\left[f(-p^0), -u p^1 - \sqrt{\Omega^2-i\epsilon}\right] \right\rbrace.
\end{align}
Noting that $f(ip^0)+f(-ip^0)$ is real, the imaginary part of the effective
action becomes:
\begin{align}
\label{eq:imaccionef}
{\rm Im}\, \Gamma_I &= \frac{- T \Sigma \pi}{2} \Im \int d^2 \pp \left\lbrace
  \Theta(u p^1 - \Omega)   \text{Res}\left[f(p^0), u p^1 - \sqrt{\Omega^2-i\epsilon}\right] + \Theta(-u p^1 - \Omega)   \text{Res}\left[f(-p^0), -u p^1 - \sqrt{\Omega^2-i\epsilon} \right] \right\rbrace \, .
\end{align}
In order to obtain a functional form from the expression above, we
evaluate the two residues involved:
\begin{align}
\label{eq:resi}
 \text{Res}\left[(\pm p^0), \pm u p^1 - \sqrt{\Omega^2-i\epsilon}\right]=g^4\frac{e^{2ia \sqrt{u^2p_1^2+\Omega^2-\pp^2-2 u p^1 \sqrt{\Omega^2-i \epsilon} }}}{u^2p_1^2+\Omega^2-\pp^2\mp 2 u p^1 \sqrt{\Omega^2-i \epsilon}} \left( \frac{1}{u^2p_1^2\mp 2 u p^1 \sqrt{\Omega^2-i \epsilon}}\right) \left( \frac{1}{-2  \sqrt{\Omega^2-i \epsilon}}\right) .
\end{align}

\begin{figure}[h!]
\includegraphics[scale=2]{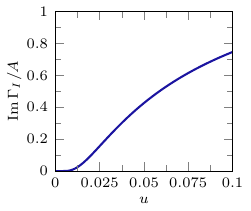}
\caption{\label{fig:numerico3+1} (Color online). Imaginary part of the effective action for thin mirrors, as a function of $u$, with $\Omega a=0.01$. $A$ is the global factor 
$A=\frac{g^4 T\Sigma (\Omega a)^3 \pi^2}{4 a^3 \Omega ^6}$. The imaginary part of the effective action, and hence the dissipative effects, are strongly suppressed for small values of the velocity between the plates.}
\end{figure}

From Eq.~\eqref{eq:imaccionef}, we see that the only modes of the vacuum
field that contribute to friction are those with
$\vert p_1\vert >\Omega/u$. Inserting Eq.(\ref{eq:resi}) into Eq.(\ref
{eq:imaccionef}), and performing the change of variables  $u p^1
\rightarrow \omega$, we obtain
%
\begin{equation}
\label{eq:funciondeomega}
{\rm Im}\, \Gamma_I =\pi T \Sigma \frac{g^4}{2\Omega} {\rm Im} \left\lbrace \int_{-\infty}^\infty dp_2  d\omega  \Theta(\omega  - \Omega)u \frac{\exp\left[2i\frac{a}{u} \sqrt{((v^2-1)\omega^2+\Omega^2u^2-p_2^2u^2-2 \omega \Omega u^2 + i \omega \epsilon u^2/ \Omega}\right]}{(u^2-1)\omega^2+\Omega^2u^2-p_2^2u^2-2 \omega \Omega u^2 + i \omega \epsilon u^2/ \Omega)\omega(\omega-2 \Omega + i \epsilon / \Omega)}\right\rbrace \, .
\end{equation}
To perform the integration over $\omega$, we use that
\begin{equation*}
\frac{1}{\omega-2\Omega+i \epsilon / \Omega}= ´{\rm p.v.} \left(\frac{1}{\omega-2\Omega}\right)-i\pi \delta(\omega-2\Omega)\, ,
\end{equation*}
and note that $\epsilon$ can be set to zero in the rest of the integrand.
Performing another change of variables, $u p_2  a \rightarrow x$, the final expression for the effective action reads
\begin{equation}
\label{eq:analiticak}
{\rm Im}\, \Gamma_I =\frac{\pi^2}{4} \frac{T \Sigma}{a^3} \frac{g^4 }{\Omega^6}(\Omega a)^4 \int_{-\infty}^{\infty} dx \frac{e^{-\frac{2}{u}\sqrt{(\Omega a)^2(4-u^2)+x^2}}}{(\Omega a)^2(4-u^2)+x^2}
\simeq \frac{\pi^2}{4} \frac{T \Sigma}{a^3} \frac{g^4 }{\Omega^6}(\Omega a)^4  \int_{-\infty}^{\infty} dx \frac{e^{-\frac{2}{u}\sqrt{4(\Omega a)^2+x^2}}}{4(\Omega a)^2+x^2}\, .
\end{equation}
This is the main result of this section, written as a product of dimensionless factors. The integral over $x$ on the Eq.\eqref{eq:analiticak} may be performed numerically. In Fig. \ref{fig:numerico3+1} we show the result for the imaginary part of the effective action as a function of $u$, for $\Omega a=0.01$. As expected, the dissipative effects are strongly suppressed as $u\rightarrow 0$. The reason behind this behavior is that there is a threshold in the energy needed to excite the internal degrees of freedom of the material, given by the frequency $\Omega$. Indeed, the integral in Eq.(\ref{eq:analiticak}) vanishes as $\exp(-4 \Omega a /u)$ for $u\ll \Omega a$,  and grows linearly in $u$
in the opposite limit. 

\begin{figure}[h!]
\includegraphics[scale=2]{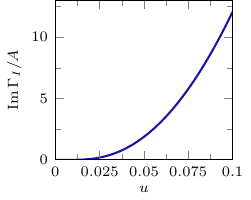}
\caption{\label{fig:numerico3+1semiesp} (Color online). Imaginary part of the effective action for half-spaces, as a function of $u$, with $\Omega a=0.01$. $A$ is the global factor $A=\frac{g^4 T\Sigma  (\Omega a)^5 \pi ^2}{16 a^3 \Omega ^8}$. The imaginary part of the effective action, and hence the dissipative effects, are strongly suppressed for small values of the velocity between the plates.}
\end{figure}

The imaginary part of the effective action for half-spaces can be obtained by integrating the thin
mirrors case, as explained above (see Eq.(\ref{eq:hat-trick2})). The result is
\begin{align}
\label{eq:accionefectivasemiespk}
{\rm Im}\, \Gamma_I &=\frac{\pi^ 2}{16}\frac{T \Sigma}{a^3} \frac{g^4}{\Omega^8} (\Omega a)^6 u^2  \int_{-\infty}^{\infty} dx \frac{e^{-\frac{2}{u}\sqrt{(\Omega a)^2(4-u^2)+x ^2}}}{\left[(\Omega a)^2(4-u^2)+x^2\right]^2}\, .
\end{align}
As already mentioned, the coupling constants $g$ for half-spaces  and thin mirrors have different dimensions.   We show the numerical results for this integration in Fig.  \ref{fig:numerico3+1semiesp},  as a function of $u$, with $\Omega a=0.01$. As in the previous case, 
there is a strong suppression for low velocities. 

%
\section{Frictional Force}\label{sec:force}
Since ${\rm Im}\, \Gamma_I >0$ when the mirrors are in relative motion,
there is an energy transfer to the system. Indeed, starting in the {\it
in} vacuum $\vert 0_{in}\rangle$, the system ends up being in an excited state, as can
be seen from the vacuum persistence probability
\begin{equation}
\vert \langle 0_{\rm out}\vert 0_{\rm in} \rangle \vert^2 = e^{-2{\rm Im} \Gamma_I }\, .
\end{equation}
Therefore, energy conservation implies that there should be force
performing mechanical work when moving the mirror. Moreover, since this
motion has a constant speed, the force has to be dissipative in nature. 
In spite of the fact that the effective action allows one to understand, in
an indirect way, the existence of a frictional force, it is not the
appropriate tool to find that force explicitly. 

On the other hand, we believe that it is important, as a consistency check,
to have an explicit, independent evaluation of that frictional force. 
To find that expression, we compute the mean value of the energy-momentum
tensor $t_{\mu\nu}$ in the in-vacuum, in the steady regime:
\begin{equation}
\langle t_{\mu\nu} \rangle \equiv \langle 0_{\rm in}\vert t_{\mu\nu}\vert 0_{\rm in}\rangle \, .
\end{equation}
The force per unit area, $\sigma$ can be obtained by means of the
point-splitting technique:
\begin{equation}
\sigma = \lim_{x\to a^+}\langle  t_{13}(x) \rangle -\lim_{x\to a^-}\langle  t_{13}(x) \rangle \, ,
\end{equation}
where 
\begin{equation}
\langle t_{13}(x)\rangle =\lim_{x'\rightarrow x} \langle \partial_1 \phi(x) \partial_3^\prime \phi(x')\rangle = \frac{1}{2} \lim_{x'\rightarrow x}  \int \frac{dp^0}{2\pi} \frac{d^2 p_\parallel}{(2\pi)^2} (i p_1)  \partial _3^\prime G_1(p^0,p_\parallel,x_3,x_3') \, .
\end{equation}
Here, $G_1$ denotes Hadamard's two-point function, that is defined by:
\begin{equation}
G_1(x,x')= \langle 0_{\rm in} \vert \lbrace \phi(x), \phi(x') \rbrace \vert 0_{\rm in} \rangle \, .
\end{equation}

Note that the very fact that there is a non-vanishing imaginary part in
$\Gamma$ implies that the in-vacuum is different from the out-vacuum; thus, in
order to compute the mean value of the energy momentum tensor one cannot
use the in-out formalism. It is well-known, however, that one can use the 
the Schwinger-Keldysh,  CTP, or
{in-in} approach \cite{CTP}.  We note that this point becomes irrelevant when
computing the static Casimir force, since when $u=0$ the two vacua are
equivalent.

In the usual {in-out} formalism, the Feynman propagator in the presence of
the mirrors can be computed perturbatively, assuming that the potentials
$V_R$ and $V_L$ are small perturbations to the free problem. We have,
schematically,    
\begin{equation}
\label{eq:diagramasinout}
G_F=G^{(0)}_F +  G^{(0)}_{F} V_{L} G^{(0)}_{F} V_{R} G^{(0)}_{F} + L \leftrightarrow R \, ,
\end{equation}
where we only included terms with mixed contributions from the L and R mirrors. We also omitted the 
integrations in the contraction of the propagators.  

In the CTP formalism, the  
free propagator is a $2\times 2$ matrix  with elements $G^{(0)}_{\alpha \beta}$, where $\alpha, \beta = +, -$: \cite{CTP}
\begin{equation}
 G^{(0)}_{\alpha\beta}(p)\equiv
 \left(
\begin{array}{c c}
{1}/({p^2+i\epsilon}) & 2\pi \delta(p^2) \theta(-p^0) \\
2\pi \delta(p^2) \theta(p^0)  & {1}/({p^2-i\epsilon})
\end{array}
\right) \, .
\end{equation}

It is worth noting that $G^{(0)}_{++}$ is defined by:
\begin{equation}
G_{++}(x,x')= \langle 0_{\rm in} \vert T \phi(x) \phi(x') \vert 0_{\rm in} \rangle \, ,
\end{equation}
and it is related to the Hadamard's function by $G_1(x,x')=-2 {\rm Im}(G_{++}(x,x'))$.

The CTP version of the perturbative evaluation of the propagator is 
\begin{equation}
\label{eq:diagramas}
G_{++}=G^{(0)}_{++}+ G^{(0)}_{+ \alpha} V_{L,\alpha \beta} G^{(0)}_{\beta \gamma} V_{R,\gamma,\delta} G^{(0)}_{\delta, +} + L \leftrightarrow R \, ,
\end{equation}
where the potentials $V_{L,R}$ are again  $2\times 2$ matrices
\begin{equation}
\lambda(p^0)=
 \left(
\begin{array}{c c}
{1}/({(p^0)^2-\Omega^2+i\epsilon}) & -\frac{\pi}{\Omega} \delta(p^0+\Omega) \\
-\frac{\pi}{\Omega} \delta(p^0-\Omega)  & {1}/({(p^0)^2-\Omega^2-i\epsilon})
\end{array}
\right) \, .
\end{equation}

Computing explicitly every contraction in Eq.\eqref{eq:diagramas}, the desired component of the energy-momentum tensor may be written as:
\begin{equation}
\label{eq:tmunupos}
\langle t_{13}(x) \rangle= - \, {\rm Im} \left\lbrace \lim_{x \rightarrow
x'}  \left[\partial_1 \partial_3^\prime G^{(0)}_{++} (x,x') + \int du dv dy
dz \partial_1 G^{(0)}_{+ \alpha} (x,u) V_{L,\alpha \beta} (u,v)
G^{(0)}_{\beta \gamma} V_{R, \gamma \delta}(y,z) \partial_3^\prime
G^{(0)}_{\delta +} (z, x') \right] \right\rbrace \, ,
\end{equation}
where we have written explicitly the spatial integrations.  The dissipative
force is given by the discontinuity of the previous magnitude at $x=a$. The
first term in Eq.\eqref{eq:tmunupos}, the contribution of the free vacuum
field propagator, is continuous at $x=a$ and will not contribute to the
force. Writing the free propagators in momentum space, the derivatives can
easily be calculated, and it can be shown that the only non-vanishing
contribution to the force comes from the term with $\delta=+$, for the
$\delta=-$ is continuous at $x=a$. The force is then given by:
\begin{equation}
\label{eq:sigma}
\sigma =  {\rm Im} \int \frac{dp^0}{2\pi} \frac{d^2 p_\parallel}{(2\pi)^2}
i p_1 G^{(0)}_{+ \alpha}(p^0,p_\parallel,a) V_{L,\alpha \beta}(p)
G^{(0)}_{\beta \gamma}(p^0,p_\parallel,a) V_{R,\gamma +}(p) + L
\leftrightarrow R \, .
\end{equation}

\begin{figure}[h!]
\includegraphics[scale=2]{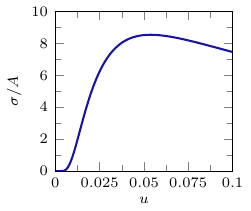}
\caption{\label{fig:fuerza3mas1} (Color online). Modulus of the dissipative force, as a function of the relative velocity of the plates, for $\Omega a = 0.01$. The global factor is $A=\frac{g^4 (\Omega a)^5}{4 \Omega ^6 \pi a^4}$. The frictional force is practically negligible for small values of the velocity between the plates.}
\end{figure}

The integrand consists of eight different terms, but only one of them turns
out to be non-vanishing: the one with $\alpha=+, \beta = \gamma = -$. The
other seven terms vanish either due to parity considerations, or as a
result of the Heaviside and Dirac delta functions appearing on the
propagators and potentials. The remaining term can easily be calculated
since the integration over $p^0$ and $p_1$ is trivial thanks to the Dirac
and Heaviside functions. The final result is 
\begin{equation}
\label{eq:fuerza}
\sigma = -\frac{1}{4\pi  a^4} \frac{g^4}{\Omega^6}(\Omega a)^6\frac{1}{u}
\int_{-\infty}^{\infty} dx \frac{e^{-\frac{2}{u}\sqrt{(\Omega
a)^2(4-u^2)+x^2}}}{(\Omega a)^2(4-u^2)+x^2} \;.
\end{equation}
This is the main result of this section, which we have chosen to write extracting a 
$1/a^4$ factor, which yields the proper dimensions dimensions to $\sigma$, times dimensionless
factors.
Note that we are left with the same integral  that we found while
calculating the effective action; this should not be surprising, since both
quantities account for the dissipative effects present on the system. This
expression can be numerically integrated, and the result is shown in Fig.
\ref{fig:fuerza3mas1} .

The behaviour of the force as a function of the relative velocity of the
plates shows that, as expected, quantum friction is practically 
negligible for small velocities, namely, such that $u\ll \Omega a$. It
reaches a maximum at a certain $u=u_0$ which a numerical study suggests is 
proportional to $\Omega a$ the only quantity with the dimensions of a velocity,
that can be built in terms of the parameters of the system (at this order). 

The expression obtained above, (\ref{eq:fuerza}) for the frictional force is a second
manifestation (the first one was the imaginary part of the effective
action) of the very same phenomenon, namely, dissipation by Casimir
friction. On the other hand, whenever there is dissipation one should
expect the presence of a flux of energy between different parts of the system, and
indeed between the system and its environment. 
This flux, or transfer of energy cannot, however, be derived from the
knowledge of the force, or from the imaginary part of $\Gamma$. 
However, we can proceed as follows: Firstly, the existence of a frictional
force produced by the vacuum field on the $R$ plate, which nevertheless
moves at a constant velocity $u$, implies that there must be an external
force of equal magnitude and opposite direction which does work on the
system, at a rate (power) per unit area: 
\begin{equation}
\rho_{\rm in} \;=\; \sigma \, u \;.
\end{equation}
This (externally provided) power clearly enters the system at the spatial region occupied by
$R$. This is not the end of the story as we can calculate the power that
{\em leaves\/} each surface, obtained from the flux of $\langle
t_{0i}\rangle$ on a (closed) surface infinitesimally close to either $L$ or
$R$, using the external normal to each one. In both cases, the calculation is entirely analogous to the one performed 
for the force, but changing the factor $i p_1 \rightarrow -i p_0$. As a result of this change, this calculation will differ by a factor $1/2$ from the one derived from Eq. \eqref{eq:sigma}.

Doing that for a surface infinitesimally close to $L$ yields:
\begin{equation}
\lim_{x\to 0^+}\langle  t_{03}(x) \rangle -\lim_{x\to 0^-}\langle
t_{03}(x) \rangle \,=\, - \frac{1}{2} \, \sigma \, u \,\equiv\, - \rho_{\rm L}
\;,
\end{equation}
in other words, a power $\rho_{\rm L} = \frac{1}{2} \rho_{\rm in}$  {\em enters\/}
the $L$ surface. Note that there is no external energy entering the system
at $L$. 

Finally, evaluating the discontinuity of the same object at $R$, we see
that:
\begin{equation}
\lim_{x\to a^+}\langle  t_{03}(x) \rangle -\lim_{x\to a^-}\langle
t_{03}(x) \rangle \,=\, \frac{1}{2} \, \sigma \, u \,,
\end{equation}
which differs from the one for $L$ just in sign. The balance of energy at
$R$ then requires to take into account the power $\sigma u$ that enters the system at
$R$, and the $\frac{1}{2}\sigma u$  part that leaves it. Thus the net power
$\rho_R$ dissipated at $R$ equals the one at $L$, and we have the equation
for the balance of power:
\begin{equation}
\rho_{\rm in} \,=\, \rho_{\rm L} + \rho_{\rm R} \;,
\end{equation}
with $\rho_{\rm L} = \rho_{\rm R}$. 

\begin{figure}[h!]
\includegraphics[scale=2]{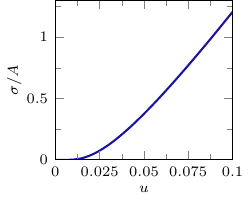}
\caption{\label{fig:fuerzasemi} (Color online). Modulus of the dissipative force for half-spaces, as a function of the relative velocity of the plates, for $\Omega a = 0.01$. The global factor is $A=\frac{g^4 (\Omega a)^5}{16 \Omega ^8 \pi a^4}$. The frictional force is practically negligible for small values of the velocity between the plates.}
\end{figure}

It is a matter of performing an integration to find the force
corresponding to half-spaces, since the same argument used for the
effective action applies here. Thus, for two half-spaces, we have:
\begin{eqnarray}
\label{eq:fuerza2}
\sigma &=& - \frac{g^4}{4 \pi \Omega u} \int_a^\infty ds
(s-a) \,\int dk
\frac{\exp[-\frac{2s}{u}\sqrt{\Omega^2(4-u^2)+k^2}]}{\Omega^2(4-u^2)+k^2}\nonumber\\
&=& - \frac{1}{16\pi a^4}\frac{g^4}{\Omega^8}(\Omega a)^7 u \int dx
\frac{\exp[-\frac{2}{u}\sqrt{(\Omega a)^2(4-u^2)+x^2}]}{[(\Omega a)^2(4-u^2)+x^2]^2}\; ,
\end{eqnarray}
a result which is shown in Figure~\ref{fig:fuerzasemi}. We recall that the
mass dimensions of $g$ and $\Omega$ are different when the mirrors have a
non-zero width, as explained in \ref{ssec:micro}. Besides, note that there
is no maximum for the force, as it was the case for the zero width plates.
The reason for that difference is that, at this order, the result for
half-spaces is obtained by integration of the zero-width case. Moreover,
the integration includes a weight factor which favours distances bigger
than $a$. It can be seen that the two half-spaces may be approximately
described as two zero-width ones but at an effective, velocity-dependent
distance. This effect appears to be responsible for the washing out the
peak that exists for two zero-width plates.

\section{Conclusions}\label{sec:concl}

In this paper we have used a functional approach to study quantum 
friction effects on imperfect moving mirrors, within a model where a 
scalar vacuum field is coupled to microscopic degrees of freedom confined 
to two mirrors, moving with a constant relative speed.  This coupling induces, 
after integrating out the microscopic degrees of freedom, a nonlocal interaction 
term in the action for the vacuum field, having an structure which depends on 
the relative velocity between the mirrors. The nonlocal action for the vacuum 
field has been used to approach the problem from two complementary viewpoints.

In the first part of the paper, we computed the imaginary part of the {in-out} effective action.
Being related to the vacuum persistence amplitude, the presence of an imaginary part is a signal 
for dissipative effects. In order to clarify the relation 
between dissipation and the analytic properties
of the nonlocal interaction (i.e. of the analogue of the dielectric permittivity), 
we performed a detailed analysis  in real time. 
The integration of the microscopic degrees of
freedom and of the vacuum field in the {in-out} functional integral 
involve the $-i\epsilon$ prescription and therefore the presence of
the Feynman propagator, whose analytic structure determines
both the analytic structure of the dielectric permittivity and of 
 the full effective action,
implying the existence
of quantum friction effects. This feature has already been observed in \cite{debate}, where the presence of singularities had been pointed out as the source of the dissipation. The analysis of the present paper 
complements the Euclidean  approach  of Ref.\cite{Fosco2011} and 
clarifies the issue of the validity of the Wick rotation of the Euclidean results.

In the second part of the paper, we computed the frictional force between mirrors. In order to do this,
we used the CTP formalism, which is crucial to obtain the correct result 
for $\langle 0_{\rm in}\vert t_{\mu\nu}\vert 0_{\rm in}\rangle$.
The crucial point here is that, due to dissipation, the {in} and
{out} vacuum states are different. This is the reason why the CTP formalism is not required to compute
static Casimir forces, while its use is unavoidable to compute the force on moving mirrors \cite{Fosco:2007nz}.

The approach described here can be generalized to the more realistic case
of the electromagnetic field at non vanishing temperature. The nonlocal
interaction should be generalized accordingly, and will involve the
derivatives of the potential vector $A_\mu$ on the position of the mirrors.
Work in this direction is in progress.

\section*{Acknowledgements}
This work was supported by ANPCyT, CONICET, UBA and UNCuyo.

\end{document}